\documentclass[aps,pre,twocolumn,floatfix]{revtex4}
\usepackage[utf8]{inputenc}
\usepackage{amsmath}
\usepackage{graphicx}
\usepackage{amssymb}
\usepackage{epsfig}
\usepackage{natbib}

\newcommand{\BE}{\begin{equation}}
\newcommand{\EE}{\end{equation}}

\usepackage{graphicx}
\usepackage{dcolumn}
\usepackage{bm}
\def\bq{\begin{equation}}
\def\eq{\end{equation}}

\begin{document}


\title{Local and nonlocal parallel  heat transport in \\ general magnetic fields}

\author{D. del-Castillo-Negrete}
\email{delcastillod@ornl.gov}
\affiliation{Oak Ridge National Laboratory \\ Oak Ridge TN, 37831-8071}
\author{L. Chac\'on}
\affiliation{Oak Ridge National Laboratory \\ Oak Ridge TN, 37831-8071}
\date{\today}

\begin{abstract}
A novel approach that enables the study of parallel transport in magnetized
plasmas is presented. The method applies to general magnetic fields with local or nonlocal parallel closures. Temperature flattening in 
magnetic islands is accurately computed.  For a wave number $k$, the fattening time scales as 
$\chi_{\parallel} \tau \sim k^{-\alpha}$ where $\chi$ is the parallel diffusivity, and $\alpha=1$ ($\alpha=2$) for non-local (local) transport. 
The fractal structure of the devil staircase temperature radial profile in weakly chaotic fields is resolved. In fully chaotic fields, the  temperature exhibits self-similar evolution of the form  $T=(\chi_{\parallel} t)^{-\gamma/2}  L \left[ (\chi_{\parallel} t)^{-\gamma/2} \delta \psi \right]$, where $\delta \psi$ is a radial coordinate. In the local case, $f$ is Gaussian and the scaling is sub-diffusive, 
$\gamma=1/2$. In the non-local case, $f$ decays algebraically, $L (\eta) \sim \eta^{-3}$, and the scaling is diffusive, $\gamma=1$.
\end{abstract}

\pacs{52.25.Fi,52.65.-y,52.25.Xz,02.70.-c}



\maketitle

The study of transport in magnetized plasmas is a problem of
fundamental interest in controlled fusion, space plasmas, and
astrophysics research. Three issues make this problem particularly
challenging: (i) The {\em extreme anisotropy} between the parallel
(i.e., along the magnetic field), $\chi_\parallel$, and the
perpendicular, $\chi_\perp$, conductivities
($\chi_\parallel/\chi_\perp$ may exceed $10^{10}$ in fusion plasmas);
(ii) Magnetic {\em field lines chaos} which in general complicates
(and may preclude) the construction of magnetic field line
coordinates; and (iii) {\em Nonlocal parallel transport} in the limit
of small collisionality.  As a result of these challenges, standard finite difference and finite elements numerical methods 
suffer from a number of ailments.
Chief among them are the pollution of perpendicular dynamics due to
truncation errors in the discrete representation of the parallel heat
flux, and the lack of a discrete maximum principle (to enforce temperature
positivity).  

Despite the severity of these issues, recent studies have succeeded in
partially addressing some of them, and important progress has been made in the study of parallel transport. 
Reference~\cite{held_etal_2004} discussed a finite element numerical implementation of nonlocal heat transport with applications including temperature flattening across magnetic islands and tokamak disruptions. 
The use of  high-order discretizations has been shown to
mitigate numerical pollution of the perpendicular dynamics in
finite-difference \cite{guenter-jcp-05-anis,guenter-jcp-07-anis}
and finite-element methods \cite{guenter-jcp-07-anis,nimrod2}.
A maximum principle has been shown to be enforceable by the use of limiters at the
discrete level in finite differences
\cite{sharma-jcp-07-anis}, and finite elements
\cite{kuzmin-jcp-09-anis}, albeit with the effect
of rendering both spatial discretizations formally first-order
accurate.
In Ref.~\cite{hudson_2008} a second order finite-difference iterative Krylov method was used to find the steady state solution of the heat transport equation in a weakly chaotic magnetic field. 

Motivated by the strong anisotropy typically encountered in magnetized plasmas ($\chi_\parallel/\chi_\perp \sim 10^{10}$), we study parallel heat transport in the extreme anisotropic regime ($\chi_{\parallel}/\chi_\perp \rightarrow \infty$).  
In addition to the previously mentioned numerical difficulties, this  regime presents potentially insurmountable issues regarding the algorithmic inversion of the discretized transport equation. In particular, the potentially degenerate null space of the parallel transport operator might render the use of state-of-the-art scalable iterative inversion methods (e.g. multigrid methods) impractical. 
To overcome these  numerical challenges, we present a novel Lagrangian Green's function approach. The proposed method  bypasses the need to discretize and invert the transport operators on a grid and allows the integration of the parallel transport equation without perpendicular pollution, while preserving the positivity of the temperature field. The method is applicable to local  and non-local  transport in integrable, weakly chaotic, and fully chaotic magnetic fields. 

Beyond the integration method, this Letter presents novel physics results. The difference between local and non-local  parallel temperature mixing and flattening inside magnetic islands is studied. Of particular interest is the dependence of the temperature relaxation time on the wave number of the temperature perturbation. Understanding this problem is of significant interest to assess the impact of magnetic islands on magnetic  confinement. 
Parallel transport is also studied in the case of weakly chaotic and fully  chaotic $3$-D magnetic fields. In the case of weakly chaotic fields, the fractal Hamiltonian structure of the magnetic islands implies that, as the ratio $\chi_{\parallel}/\chi_\perp$ increases,  the radial temperature profile approaches a devil-staircase. Going beyond previous studies \cite{hudson_2008}, 
we unveil the fractal structure of the devil-staircase in the previously unaccessible $\chi_\perp = 0$ regime.
These results open the possibility of a deeper understanding of the role of cantori which have been observed to act as partial transport barriers in numerical studies \cite{hudson_2008} and experiments 
\cite{rfp_2009}.  Another problem of considerable interest in fusion and astrophysical plasmas is the understanding of electron heat transport in fully chaotic fields. The case of local parallel transport  has been extensively studied since the pioneering work in  Refs.~\cite{jokipii_1968,rechester_rosenbluth_1978}. However, most studies have restricted  their attention to local transport. 
Here, in addition of the study of the local transport, we present novel results on the self-similar, non-Gaussian spatio-temporal scaling of radial 
heat transport in the non-local case. 
 
Our starting point
is the heat transport equation in a constant density plasma \bq
\label{eq_1}
\partial_t T = -\nabla \cdot {\bf q} \, ,
\eq
where ${\bf q}$ is the heat flux. 
For local transport, in the limit  
$\chi_\perp=0$, 
\bq
\label{eq_2}
{\bf q}= - \chi_{\parallel}  \left[  \hat{ \bf b} \cdot \nabla T  \right ]   \hat{\bf b}\, ,
\eq
where  $\hat{\bf b} ={\bf B}/|B|$. Substituting (\ref{eq_2}) into  (\ref{eq_1}), we get 
\bq
\label{eq_3}
\partial_t T = -\partial_s q_{\parallel} \, , \qquad
q_{\parallel}=-\chi_\parallel \partial_s T \, , \eq where
$\partial_s=\hat{ \bf b} \cdot \nabla$ is the derivative along
the field line, and we have assumed the tokamak ordering
${\partial_s \ln B} \approx 0$.
In the case of non-local transport, following Ref.~\cite{del_castillo_2006}, we consider
\bq
\label{eq_5}
q_{\parallel}=\frac{\lambda \chi}{\pi} \int_0^\infty
\frac{T\left(s+z\right)-T\left(s-z\right)}{z^\alpha} dz  \, ,
\eq 
with $1 \leq \alpha \leq 2$. 
The case $\alpha =1$ reduces to the collisionless heat flux 
characterized by the free-streaming of electrons along magnetic field lines \cite{hammet_perkins_1990,held_etal_2001}, and  the case $\alpha =2$ reduces to the local diffusive case. The model in Eq.~(\ref{eq_5}) allows the interpolation between these two regimes. In particular, Eq.~(\ref{eq_5}) allows the incorporation of  parallel transport processes  with underlying non-Gaussian (Levy $\alpha$-stable) stochastic processes \cite{del_castillo_2006}.

\begin{figure}
\includegraphics[scale=0.5]{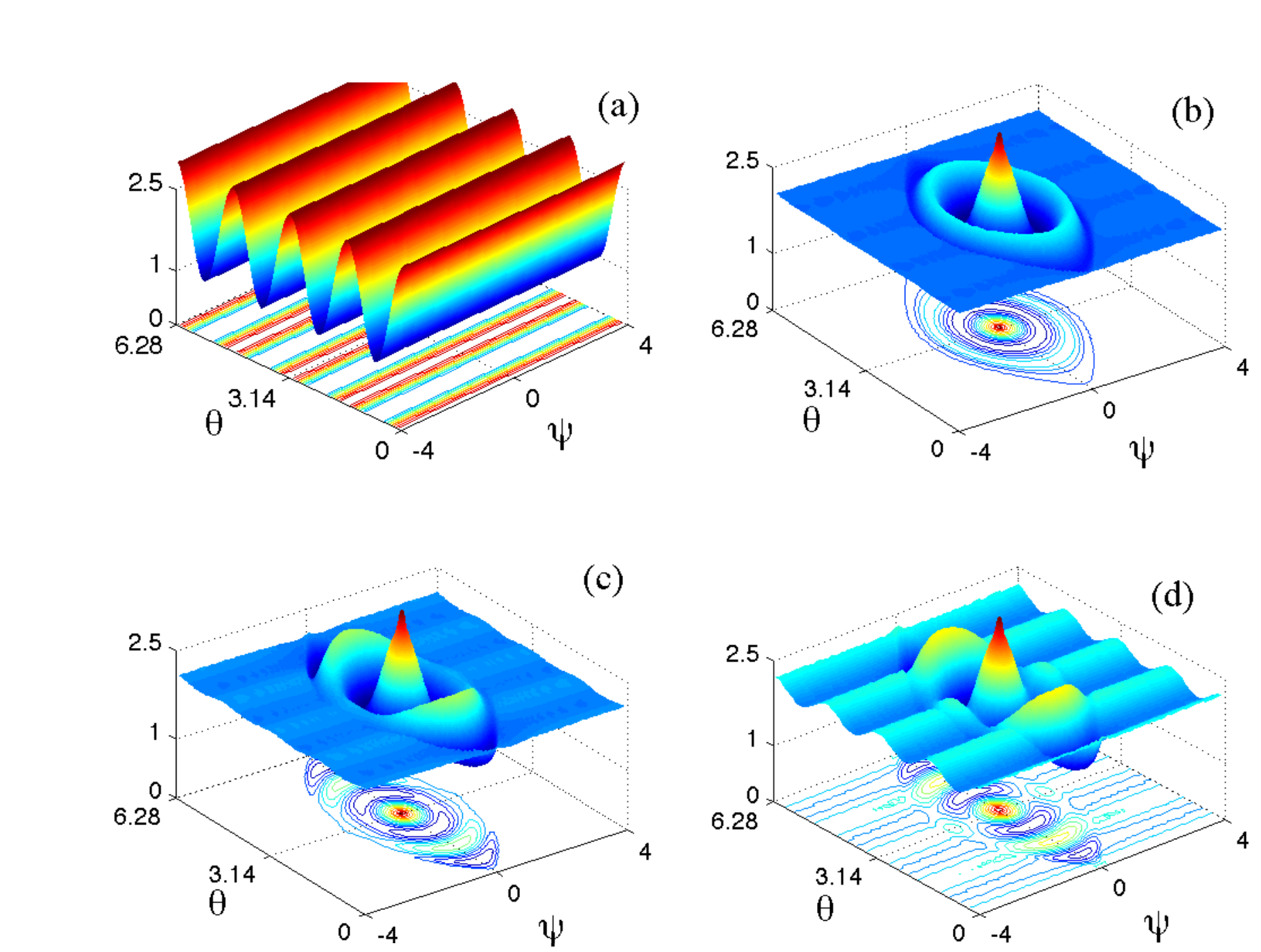}
\caption{
\label{fig_1}
Temperature mixing and homogenization in a magnetic island. Panel (a) shows the initial condition $T_0=2+\cos(4 \theta)$, and panel $(b)$ the final relaxed state, $T_f=F(A_z)$, where $A_z$ is the magnetic potential, for both local and non-local transport. Panels (c) and (d) show the solution at $\chi_{\parallel} t= 0.6$ in the local and non-local cases respectively. 
}
\end{figure}

Substituting Eq.~(\ref{eq_5}) into  Eq.~(\ref{eq_1}), and assuming $\chi_\parallel$ constant, the 
 transport equation can be written in a compact form as 
 \bq
\label{eq_6}
\partial_t T = \chi_\parallel \partial^\alpha_{|s|} T \, , 
\eq where
the operator $\partial^\alpha_{|s|}$ denotes the symmetric fractional
derivative of order $\alpha$, defined as 
${\cal F}\left[ \partial_{|s|}^\alpha T \right ]=-|k|^\alpha  \hat{T}$.
Here, ${\cal F}$ denotes the Fourier transform \cite{del_castillo_2006}. 
As expected, in the limit $\alpha=2$,
Eq.~(\ref{eq_6}) reduces to the diffusion equation in Eq.~(\ref{eq_3}).

The proposed method is based on the Green's function solution of Eq.~(\ref{eq_6}) along the Lagrangian trajectories of the magnetic field.  
The unique  magnetic field line trajectory (parametrized by the arc length)  that  goes through a point 
${\bf r}_p$ is given  by the solution of 
\bq
\label{eq_9}
\frac{d {\bf r}}{d s}=\hat{\bf b} \, , \qquad {\bf r}(s=0)={\bf r}_p \, .
\eq
Thus, given an initial condition in the whole domain, $T({\bf r},t=0)$, the initial condition along the field line is 
$T_0(s)=T({\bf r}(s),t=0)$ and the temperature at ${\bf r}={\bf r}_p$ at time $t$ is  given by 
\bq
\label{eq_8}
T({\bf r}_p,t) = \int_{s_1}^{s_2} T_0 \left[ {\bf r}(s) \right]
G_\alpha(s,t) ds\, , \eq 
where $G_\alpha$ is the Green's function of Eq.~(\ref{eq_6}). 
For unbounded field lines, $(s_1,s_2)=(-\infty, \infty)$ and the Green's function is given by
\bq
\label{greens}
G_\alpha(s,t)=\frac{1}{2 \pi} \int_{-\infty}^{\infty} e^{-  \chi t \left| k \right |^\alpha - i k s} d k \, .
\eq
For $\alpha=2$, Eq.~(\ref{greens})  gives the Gaussian distribution
\bq
\label{eq_10}
G_2(s,t) =\frac{1}{2 \sqrt{\pi}}\left( \chi_\parallel t\right)^{-1/2} \exp\left[ 
-\frac{s^2}{4 \chi_\parallel t} \right] \, ,
\eq
and in the non-local free streaming case, $\alpha=1$, it gives  the Cauchy distribution
\bq
\label{eq_11}
G_1(s,t) = \frac{\left( \chi_\parallel t 
\right)^{-1}}{\pi}\frac{1}{1+\left(s/\chi_\parallel t\right)^2}\, .
\eq
For general $\alpha$, $G_\alpha=(\chi_\parallel t)^{-1/\alpha} L_{\alpha,0}[ (\chi_\parallel t)^{-1/\alpha} s]$ where $L_{\alpha,0}$ is the symmetric $\alpha$-stable Levy distribution, see for example Ref.~\cite{del_castillo_2004}. 
In the case of closed (periodic) field lines, the integration domain in Eq.~(\ref{eq_8})  is a finite interval 
$(s_1,s_2)$ with ${\bf r}(s_1)={\bf r}(s_2)$, and one has to use the periodic Green's function, $G^P_\alpha$, obtained by mapping the unbounded Green's function, $G_\alpha$, into the periodic domain.

\begin{figure}
\includegraphics[scale=0.28]{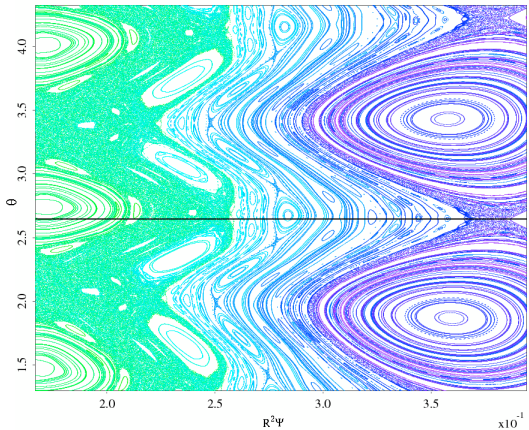}
\caption{
\label{fig_2}
Poincare plot of the weakly chaotic magnetic field used in solution of the parallel heat transport equation shown in Fig.~\ref{fig_3}.     
}
\end{figure}

\begin{figure}
\includegraphics[scale=0.5]{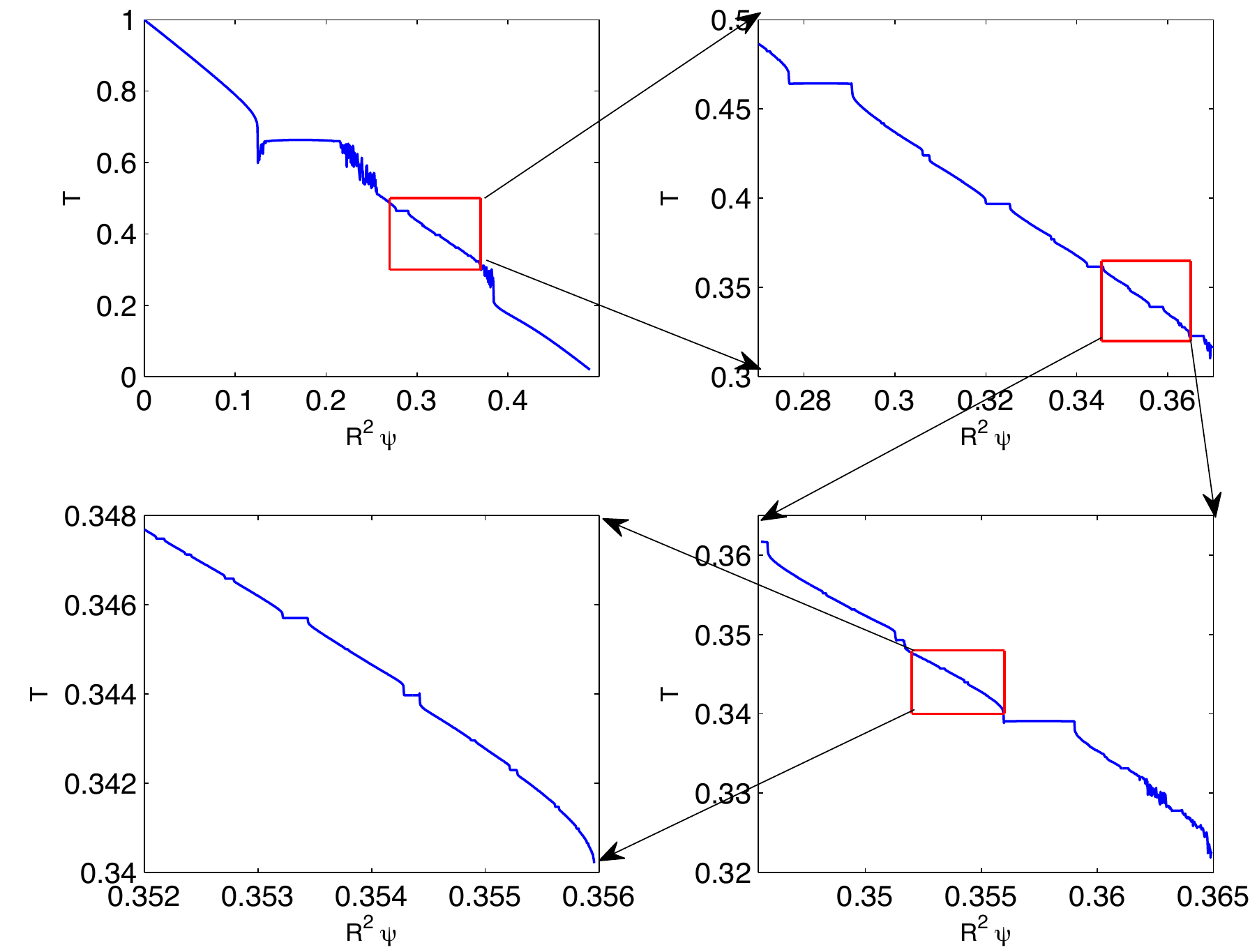}
\caption{
\label{fig_3}
Radial temperature  profile of the time-asymptotic solution of the parallel heat transport equation for the weakly chaotic magnetic field in Fig.~\ref{fig_2}. The zooms in the successive panels unveil the fractal structure of the devil-staircase profile.   
}
\end{figure}

At this point it is important to indicate a
fundamental difference between the work in Refs~\cite{held_etal_2001,held_etal_2004} and the method proposed here. 
In Refs.~\cite{held_etal_2001,held_etal_2004}, the flux is calculated throughout the computational domain by integrating along the field lines. However, in these references, once the flux is computed, the Lagrangian approach is abandoned and the flux is mapped to Gaussian quadrature points for the finite-element standard integration of the temperature evolution equation on a grid. On the other hand, the method proposed here is fully Lagrangian and completely bypasses the use of finite differences or finite elements integration schemes,  circumventing the numerical limitations discussed before. Another clear advantage of the use of the Green's function is that the solution at an arbitrary  time $t=t_f$ is obtained directly from the integral in Eq.~(\ref{eq_8}). This is to be contrasted with the standard explicit and implicit time discretization schemes that require the knowledge of  the solution at previous times $t<t_f$ before  the solution at $t=t_f$ can be constructed. 
A similar decoupling occurs in space. 
According to Eq.~(\ref{eq_8}), the evaluation of $T$ at  ${\bf r}_p$ is fully decoupled from the solution  at 
${\bf r} \neq {\bf r}_p$, yielding naturally to a massively parallel approach.  


In what follows, we use the Lagrangian Green's function method 
to study local and non-local parallel transport in magnetic fields of increasing levels of complexity. 
We start with the study of transport in magnetic islands, which we model with a $2$-D magnetic field with vector potential 
$A_z=\frac{1}{2} \psi^2+\epsilon \cos \theta$, where $\psi$ and $\theta$ are radial and angular coordinates respectively. Of particular interest is the difference in the temperature relaxation properties  in the presence of   local and nonlocal transport. 
Figure~1 shows the solution of the parallel transport of an initial condition of the form $T_0=2+\cos(k \theta)$ with $k=4$. As expected, for large enough times, the temperature relaxes to a unique solution of the form $T_f=F(A_z)$ in both the local and non-local cases. However,  the dynamic process leading to the final relaxed state is different. As the figure shows, temperature mixing in the non-local case 
is less homogeneous and slower than in the local case. In general, from Eq.~(\ref{greens}) it follows that the homogenization time of a temperature perturbation with mode number $k$ scales as  $\chi_\parallel \tau \sim k^{-\alpha}$ with $\alpha=2$ ($\alpha=1$) in the case of local (non-local)  transport. 

For the study of transport in $3$-D magnetic fields, we assume cylindrical geometry periodic in $z$
with period $L=2 \pi R$ where $R$ is a  
constant.  The  magnetic field consists of a perturbed screw-pinch of the form
\bq
{\bf B}= (r B /\lambda)/[1+(r/\lambda)^2] \hat{\bf e}_\theta + B_0 \hat{\bf e}_z + {\bf B}_1(r,\theta,z) \, ,
\eq
with $B$, $\lambda$, and $B_0$ constants. The magnetic potential of the perturbation, ${\bf B}_1=\nabla \times A_z \hat{\bf e}_z$, consists of a superposition of modes 
\bq
\label{eq_20}
A_z(r,\theta,z)=\sum_{m,n} A_{mn}(r)  \cos \left( m \theta - n z/R + \zeta_{mn} \right ) \, ,
\eq
with
\bq
\label{eq_21}
A_{mn}= \epsilon a(r) \left( \frac{r}{r_*} \right)^m \exp \left[ 
\left( \frac{r_*-r_0}{\sqrt{2} \sigma}\right)^2 
-\left( \frac{r-r_0}{\sqrt{2} \sigma}\right)^2  \right] \, .
\eq
For each  $(m,n)$, the values of $r_*$and $r_0$ are chosen so that the safety factor satisfies $q(r_*)=m/n$ and $d A_{mn}/dr (r=r_*)=0$. The prefactor $(r/r_*)^m$ is included to guarantee the regularity of the radial eigenfunction near the origin, $r\sim 0$. The function, $a(r)=[1-\tanh[(r-1)/l]]/2$, is introduced to guarantee the vanishing of the perturbation for $r \sim 1$ and the existence of well-defined flux surfaces at the plasma boundary.

\begin{figure}
\includegraphics[scale=0.4]{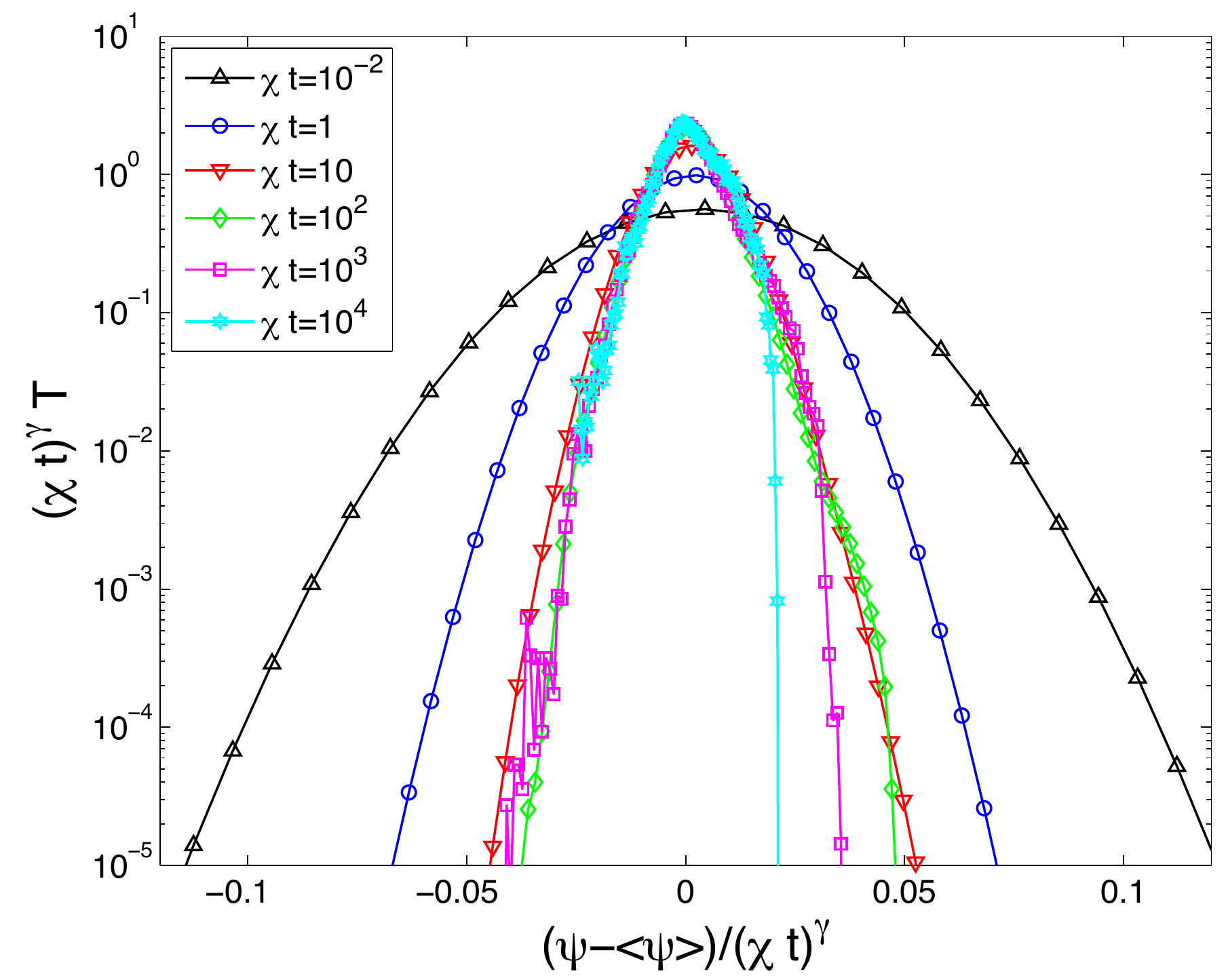}
\caption{
\label{fig_4}
Self-similar spatio-temporal evolution of the radial temperature profile for local, $\alpha=2$, transport in a fully chaotic magnetic field.   In this case the scaling function is Gaussian and the scaling exponent is $\gamma=1/2$.    
}
\end{figure}

\begin{figure}
\includegraphics[scale=0.4]{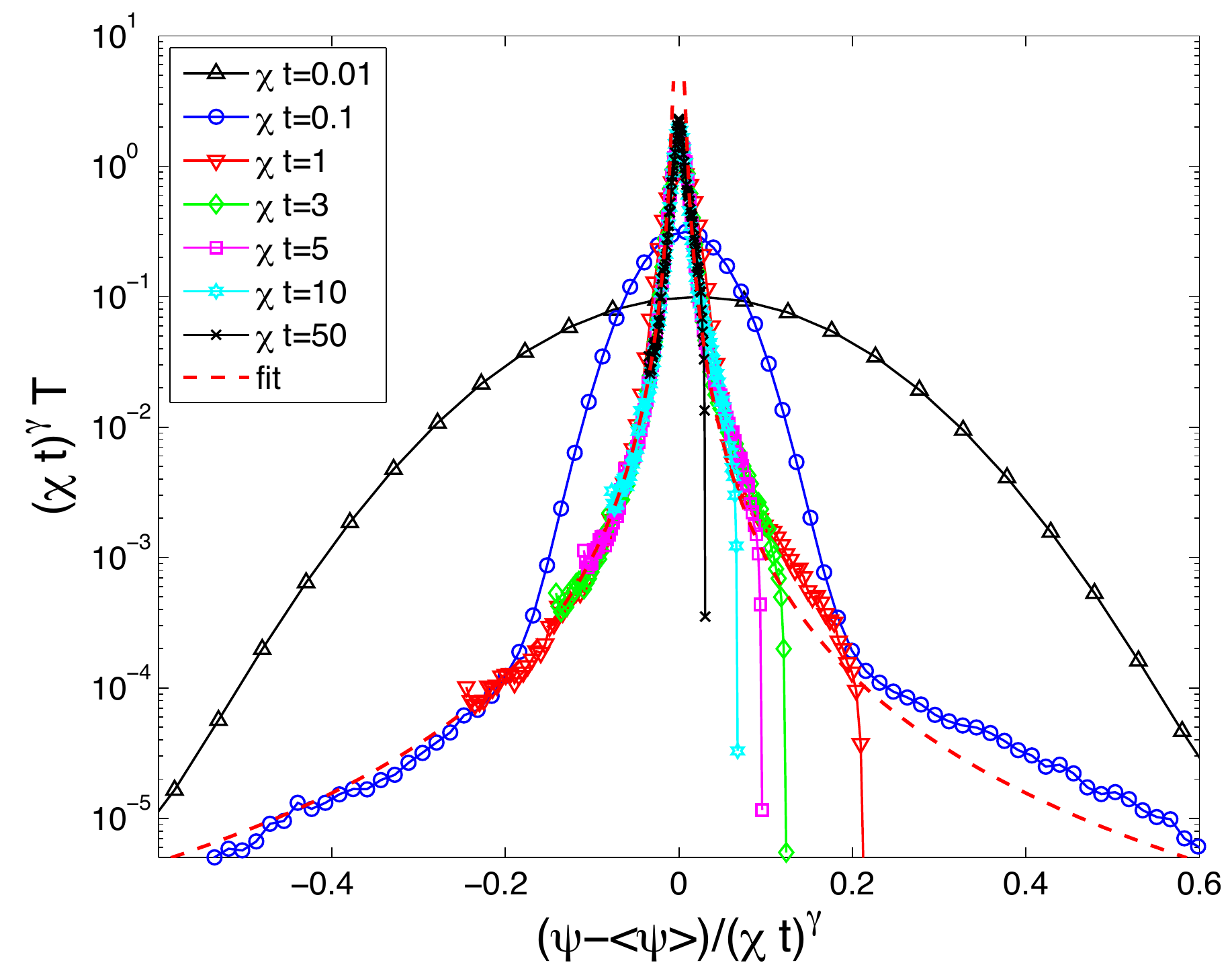}
\caption{
\label{fig_5}
Self-similar spatio-temporal evolution of the radial temperature profile for non-local, $\alpha=1$, transport in a fully chaotic magnetic field.   In this case the scaling function is strongly-non-Gaussian and, as the dashed-line fit shows,  exhibits algebraic decay of the form $L \sim \eta^{-3}$.  The scaling exponent is $\gamma=1$. 
}
\end{figure}

In the study of transport in weakly chaotic fields, only two modes were included. As the Poincare plot in
Figure~\ref{fig_2} shows, the magnetic field  in this case exhibits a rich fractal-like structure resulting from the existence of higher-order resonances. Figure~\ref{fig_3} shows the time asymptotic, final radial temperature solution along the $\theta = 2.7$ horizontal line in Fig.~\ref{fig_2}, corresponding to the initial condition $T_0=1- 2 R^2 \psi$. As expected, the same state is reached in the presence  of local and non-local transport. However, as mentioned before, the relaxation time is longer in the non-local case.  The fractal structure of the magnetic field gives rise to a devil-staircase temperature profile in which horizontal sections (resulting from high-order resonances) are mixed with vertical sections (resulting from KAM invariant circles and Cantori).

To study transport in a fully chaotic magnetic field we considered a set of $21$ strongly  overlapping modes. In this case, the Poincare plot (not shown) is fully hyperbolic and does not exhibit any structure. 
The initial condition consists of a narrow ``cylindrical shell" of the form $T_0=\exp [ -R^2 (\psi-\psi_0)/\sigma]^2$, with $\psi_0=0.18$ and $\sigma=0.05$. Figures~\ref{fig_4} and ~\ref{fig_5} show the time evolution of the radial profile of the temperature averaged in $\theta$ and $z$ in the local and the non-local free streaming cases. In both cases, the temperature exhibits an asymptotic  self-similar evolution of the form
\bq
\langle T \rangle_{\theta,z} (\psi,t) = \left( \chi t \right)^{-\gamma/2} L(\eta) \, ,
\eq
where the similarity variable is defined as  $\eta=(\psi -<\psi>)/(\chi t)^{\gamma/2}$ with $\gamma$ the scaling exponent. 
From here, it follows that the second moment scales as
$<\psi^2> \sim t^{ \gamma}$.  As  Fig.~\ref{fig_4} shows, in the local transport case the scaling function is approximately Gaussian, $L\sim e^{-\eta^2/\sigma}$. Consistent with Ref.~\cite{rechester_rosenbluth_1978}, in this case $\gamma \approx 1/2$. This subdiffusive scaling results from the combination of the diffusive transport along the field line and the 
quasilinear diffusion of the field line itself due to the magnetic field line chaos.  
On the other hand,  in the nonlocal case, the scaling function is strongly non-Gaussian and exhibits an algebraic decay of the form $L\sim \eta^{-3}$.  Interestingly, in the non-local case $\gamma \approx 1$. This diffusive scaling results from the coupling of the quasilinear diffusion of the field line due to chaos and the free-streaming transport along the field line. 

Summarizing, this Letter presented a Lagrangian Green's function method for the accurate and efficient  computation of purely parallel  ($\chi_\perp=0$) local and non-local transport  in arbitrary magnetic fields with constant  $\chi_\parallel$. 
Because of the parallel nature
of the Lagrangian calculation, the formulation naturally
leads to a massively parallel implementation, suitable for today's
supercomputers.
The method was applied to compute temperature mixing in magnetic islands, for which the ratio of the non-local free-streaming, $(\chi_{\parallel} \tau)_{nl}$, and diffusive, $(\chi_{\parallel} \tau)_{d}$,
transport relaxation times scales as $(\chi_{\parallel} \tau)_{nl}/(\chi_{\parallel} \tau)_{d} \sim k$. Radial transport in 
$3$-D magnetic fields in cylindrical geometry was also studied. In the case of weakly chaotic fields, the fractal structure of the devil staircase of the radial temperature profile was resolved. In the case of fully chaotic fields, the radial transport exhibited self-similar spatio-temporal behavior. Contrary to the well-known diffusive case, the non-local case exhibits a non-Gaussian, algebraic decaying temperature profile with scaling exponent $\gamma=1$. Current work includes the implementation  of the Lagrangian Green's function method as integral  part of a more general multiscale framework that incorporates finite $\chi_\perp$, non-constant $\chi_\parallel$, and heat sources. 

 \section{Acknowledgments}
This work was sponsored by the Oak Ridge National Laboratory, managed by UT-Battelle, LLC, for the U.S.Department of Energy under contract DE-AC05-00OR22725.


\pagebreak

\end{document}